\documentclass[iop]{emulateapj}

\shorttitle{RR Lyrae stars in the Leo V dSph}
\shortauthors{Medina et al.}

\begin{document}

\title{Serendipitous discovery of RR Lyrae stars in the Leo V ultra-faint galaxy}

\author{Gustavo E. Medina\altaffilmark{1,3}, Ricardo R. Mu\~noz\altaffilmark{1}, A. Katherina Vivas \altaffilmark{2}, Francisco F\"{o}rster\altaffilmark{3,4}, Jeffrey L. Carlin\altaffilmark{5}, Jorge Martinez\altaffilmark{1,4,3}, Lluis Galbany\altaffilmark{7}, Santiago Gonz\'alez-Gait\'an\altaffilmark{4,3},  Mario Hamuy\altaffilmark{1,3}, Thomas de Jaeger\altaffilmark{6,3,1}, Juan Carlos Maureira\altaffilmark{4}, Jaime San Mart\'in\altaffilmark{4}
}

\altaffiltext{1}{Departamento de Astronom\'ia, Universidad de Chile, Casilla 36-D, Santiago, Chile.}
\altaffiltext{2}{Cerro Tololo Inter-American Observatory, Casilla 603, La Serena, Chile.}
\altaffiltext{3}{Millenium Institute of Astrophysics MAS.}
\altaffiltext{4}{Center for Mathematical Modelling, Universidad de Chile, Av. Blanco Encalada 2120 Piso 7, Santiago, Chile.}
\altaffiltext{5}{LSST, 933 North Cherry Avenue, Tucson, AZ 85721, USA.}
\altaffiltext{6}{Department of Astronomy, University of California, Berkeley, CA 94720-3411, USA}
\altaffiltext{7}{PITT PACC, Department of Physics and Astronomy, University of Pittsburgh, Pittsburgh, PA 15260, USA}

\begin{abstract}
During the analysis of RR Lyrae stars discovered in the High cadence Transient Survey (HiTS) taken with the Dark Energy Camera at the $4-$m telescope at Cerro Tololo Inter-American Observatory, we found a group of three very distant, fundamental mode pulsator RR Lyrae (type $ab$). 
The location of these stars agrees with them belonging to the Leo V ultra-faint satellite galaxy, for which no variable stars have been reported to date. 
The heliocentric distance derived for Leo~V based on these stars is $173\pm5$\,kpc. 
The pulsational properties (amplitudes and periods) of these stars locate them within the locus of the Oosterhoff II group, similar to most other ultra-faint galaxies with known RR Lyrae stars. 
This serendipitous discovery shows that distant RR Lyrae stars may be used to search for unknown faint stellar systems in the outskirts of the Milky Way.
\end{abstract}

\keywords{Galaxy: halo - stars: variables: RR Lyrae - galaxies: individual (Leo V) - Local Group}
\maketitle

\section{Introduction}
With the advent of large, optical sky surveys like the Sloan Digital Sky Survey \citep[SDSS;][]{York2000a}, the Panoramic Survey Telescope And Rapid Response System (Pan-STARSS$-1$, \citealt{chambers16a}) or the Dark Energy Survey \citep[DES;][]{abbott15}, over the last decade and a half, a flurry of new Milky Way (MW) satellites have been discovered \citep[e.g.,][]{willman05a, belokurov06a, belokurov06b,belokurov08a, belokurov10a,zucker06a, zucker06b,irwin07a,bechtol15a,koposov15a,drlica15,drlica16,martin15}.
These discoveries are of particular relevance since they allow us to probe the faint end of the galaxy luminosity function and shed new light into known discrepancies between predictions from cosmological simulations and observations.
Among these stands the well-known ``missing satellites problem" present in $\Lambda$CDM models \citep[e.g.,][]{Kauffmann1993, klypin99a, Moore1999, Simon2007} wherein hundreds to thousands of low mass subhalos should orbit around the MW but only a few dozen actual dwarf satellites are known.
In this context, a reliable census of satellite galaxies, particularly at the faint end, is essential to make progress toward solving these inconsistencies.
This has led to focused efforts to discover dwarf galaxies and extremely low  luminosity sub-halos, and finding new ways to use the data available from wide and deep field surveys in an efficient way \citep[e.g.,][]{bechtol15a,koposov15a,Baker2015}.

The detection of variable stars in dwarf satellites has played a major role in the study of their properties and populations. 
The case of the pulsating RR Lyrae stars (RRLs) is particularly interesting since they are old stars ($>10$ Gyr) easily identifiable by the shape of their light curves. 
In fact, historically the discovery of RRLs in dwarf spheroidal (dSph) galaxies was the first confirmation that these systems contained old, population~II stars \citep[e.g.,][]{baade39a, saha86a, siegel06a}. 
Additionally, these type of variable stars play an important role since they are considered well-known standard candles and therefore provide reliable distances to their host dwarf satellites.
In this context, at least one RRL has been reported in every Milky Way's satellite classified as a dwarf galaxy that has been searched for them \citep[see compilation in][]{Vivas2016}, including systems with extremely low luminosity and surface brightness like Segue~1 \citep{Simon11a} and Segue~2 \citep{Boettcher13a}. 
This fact has brought forth the idea that RRL can actually be used to discover new stellar systems in the outer halo \citep{Sesar2014,Baker2015,Sanderson2016}, as well as to study the properties of halo substructures \citep{Vivas2001,Sesar2013,Dra13_100kpc,Torrealba2015}.
Since distant RRLs are rare, these works suggest that they can trace the existence of faint stellar systems (as the light of the lighthouse). 
In particular, \citet{Baker2015} suggested that groups of two or more RRLs at heliocentric distances $>50$\,kpc could reveal stellar systems as faint as $M_V=-3.2$.

In this Letter we describe the serendipitous discovery of variable stars in the Leo~V ultra-faint galaxy while studying a sample of distant RRLs from the High cadence Transient Survey \citep[HiTS;][]{forster16a}.
Following \citeauthor{Baker2015}'s idea, two close groups of RRLs were recognized in these data. 
The locations of these groups agreed with the locations of the dwarf satellites Leo~IV and Leo~V. 
Variable stars in Leo~IV have been identified before by \citet{Moretti2009}, but no search so far has been reported in Leo~V.

Leo~V was discovered in SDSS data by \citet{belokurov08a}. 
It is a faint system, $M_V=-4.4$ \citep{Sand2012}, composed of an old, metal-poor ([Fe/H]$=-2.48$; \citealt{Collins2016}) stellar population. 
Based on the observed horizontal branch (HB), the estimates of the distance to Leo~V have been set between $175$ to $195$\,kpc \citep{belokurov08a,deJong2010,Sand2012}.
Its closeness with Leo~IV in both location in the sky and radial velocity has suggested a possible common origin for both galaxies \citep{belokurov08a,deJong2010,Blana2012}. 
It has also been suggested that the galaxy is undergoing tidal disruption \citep{belokurov08a,deJong2010,Collins2016} although \citet{Sand2012} did not find evidence for extra-tidal features in their data. 
In any case, the detection of RRLs in this work allows the determination of a precise distance to this galaxy which will be useful for future dynamical works to understand the origin of this system and its possible interaction with Leo~IV and/or the Milky Way.

The structure of this article is as follows: In \S2 the details of the HiTS observations are presented.
Properties of the RRLs discovered in Leo~V are presented in  \S3 and finally, a summary and final discussion are addressed in  \S4.
The details of the methodology and analysis of the complete list of distant RR Lyrae stars in the HiTS survey is the topic of a separate paper (Medina et al., in preparation).

\section{Observations and Data Analysis}

The data used in this article were collected between UT 2014 February 28 and UT 2014 March 4 with the Dark Energy Camera \citep[DECam,][]{Flaugher2015}, a prime focus CCD imager installed at the Blanco $4-$m telescope at Cerro Tololo Inter-American Observatory (CTIO), as part of the High cadence Transient Survey \citep[HiTS,][hereinafter \citeauthor*{forster16a}]{forster16a}. 
The survey is focused on the detection of young supernovae with emphasis in the early stages of the explosions. 
Despite being designed for other purposes, the data can be mined for the study of optical transients in general.

HiTS observed 40 blindly selected fields in 2014 at high Galactic latitudes, covering a total of $\sim 120$ square degrees from $150^{\circ}$ to $175^{\circ}$ in RA and $-10^{\circ}$ to $3^{\circ}$ in DEC (see Figure~4 in \citeauthor*{forster16a}). 
These fields were observed on the $g$ SDSS photometric system filter with exposure times varying from $160$ seconds ($83\%$ of the total) to $174$ seconds ($14\%$), and a cadence of two hours. 
That resulted in a total of $20$ epochs for most of the fields, with a limiting apparent magnitude of $23-24.5$ (\citeauthor*{forster16a}). 
The mean seeing of the survey was $1.5$\,arcsec. 
Details on the survey strategy can be found in \citeauthor*{forster16a}, while the search for RRLs will be detailed in an upcoming paper (Medina et al. in preparation). 
Here we summarize the most relevant steps.

\begin{figure}
\includegraphics[scale=0.43]{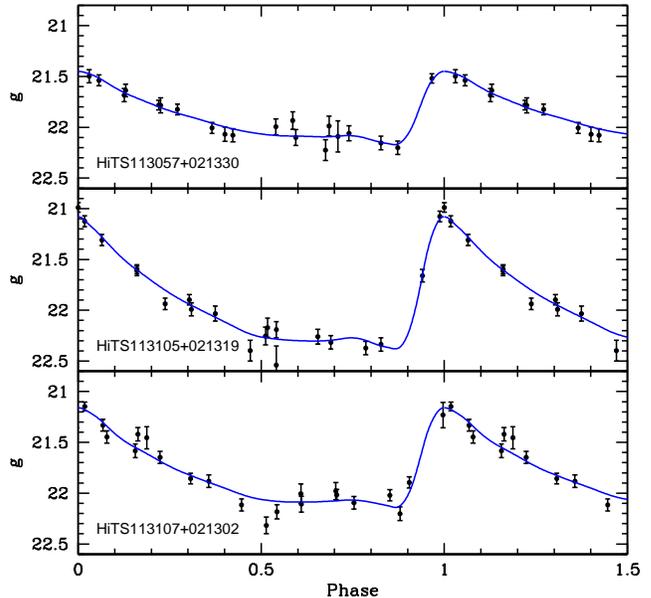}
\caption{Folded light curves for the three RRLs found in Leo~V. The solid blue line is the best fitted template from the library of \citet{Sesar2010}. 
The pulsational properties and distance of these variables are shown in Table \ref{tab:RRL}.}
\label{fig:lc}
\end{figure}

The data processing was carried out using the DECam community pipeline \citep{Valdes2014}. 
Point source photometry was done with the SExtractor photometry software \citep{Bert96}. 
To generate time series, we performed an alignment in $x,y$ position of the SExtractor outputs. 
Using our second epoch as reference, we compared the instrumental magnitudes of all stars and calculated a zero point offset ($\Delta_{\rm rel}$) on a chip-by-chip basis.
Then, the reference scan was calibrated (only by zero point) using overlapping SDSS photometry from DR10 \citep{Ahn2014}. 
To correct for extinction, we used the re-calibrated dust maps of \citet{Sch11}, and calculated the extinction values following $A_g = 3.303 \; E(B-V)$.

For the detection of RRLs, we filtered out objects with less than five observations, and also stars with low variation in brightness compared with the typical magnitude errors. 
After that we performed a period search using the Generalized Lomb-Scargle technique \citep[GLS;][]{Zech09} using an astroML python module developed by \citet{Vander12}. 
Pre-candidates were selected based on the periods found, where objects with periods shorter than $4.8$\,hours and longer than $21.6$\,hours were rejected for being outside the typical pulsation window of RRLs. 
In this process, we did not consider periods within $0.1$\,days around $0.33$ and $0.50$\,days to avoid spurious detection attributable to aliasing. 
Another filter was applied based on the level of significance of the period detection computed by the python module (statistical significance $< 0.08$ were left out). 
Following these criteria, we accepted the two most significant periods (if applicable). 
Finally, the candidates with a difference in magnitude between the brightest and faintest observation larger than $0.2$ were visually inspected to make the final catalogue. 
The pulsation parameters of the light curves were obtained by adjusting RRLs templates from SDSS Stripe 82 \citep{Sesar2010}. 
For this fitting, small variations around the observed amplitude and GLS period were allowed. 
We selected the best fit based on a $\chi^2$ minimization criterion. 
The mean magnitudes of the RRLs were calculated by integrating the transformed fitted template, in intensity units, and transforming back to magnitudes the mean.

\begin{table*}[t]\scriptsize
\caption{Identification number and main properties of the three Leo~V's RRL presented in this work.}
\label{tab:RRL}
\begin{center}
\begin{tabular}{c c c c c c c c c c c}
\hline
\hline
	ID &  R.A.  &  DEC  &  $<g>$ & $A_g$  &  $d_H$ ([Fe/H]$=-1.6$) &  $d_H$ ([Fe/H]$=-2.31$)& Period  & Amplitude  & Type  &  N\\
	 &  (deg)  &  (deg)  &    &    & (kpc) & (kpc)   & (days) & $g$ &  &  \\
\hline 
HiTS113057+021330 &	$172.73946$	&   $2.22514$  & $21.79 \pm 0.08$ & $0.09$ & $166\pm 8$ & $176\pm 9$  & $0.6453$ & $0.72$ & ab &  $20$\\\
HiTS113105+021319 &	$172.76936$	&   $2.22200$  & $21.79 \pm 0.08$ & $0.09$ & $166\pm8$ & $176\pm9$  & $0.6573$ & $1.34$ & ab &  $20$\\\
HiTS113107+021302 &	$172.77796$ &	$2.21734$  & $21.68 \pm 0.08$ & $0.09$ & $158\pm7$ & $167\pm8$  & $0.6451$ & $0.99$ & ab &  $21$\\
\hline
\end{tabular} 
\end{center}
\end{table*}

We examined the best two periods because aliasing can produce reasonable lightcurves for different periods. 
This was indeed the case for one RRL found in Leo V, HiTS113105, for which the second best period, $0.657$\,days, produced a light curve almost indistinguishable from the main period (a 1-day alias), $1.955$\,days.
We took the $0.657$\,days period as the correct one as it agrees better with the expectations for RRL stars.

\section{RR Lyrae stars in Leo~V}

Theoretical calibrations for the absolute magnitude of RRL stars in the SDSS $g-$band exist in the literature \citep{marconi06,caceres08}, but they require knowledge of color information which is not available from our single-band survey.
For the same reason, we do not have data to calculate a Johnson $V$ magnitude using known transformation equations like the ones derived by R. Lupton\footnote{\footnotesize  \url{http://www.sdss.org/dr12/algorithms/sdssUBVRITransform/ \#Lupton2005}}.
Thus, we estimated preliminary $M_g$ for our stars by comparing our complete sample of RRL with the ones in the Catalina surveys (\citealt{Dra13, Dra14}). 
Distances were computed for the RRLs in the Catalina survey using their mean $V$ magnitudes (corrected by extinction) and assuming [Fe/H]$=-1.6$ as the metallicity for the halo, which yields $<M_V>=0.55$ \citep{Dem00}. 
We found $\sim50$ stars in common between our catalogs and from the known apparent $g$ magnitudes of the stars in our sample, we determined an average absolute $g$ magnitude of $<M_g>=0.69\pm 0.06$. 
We validated this method by comparing the results for a sub-sample of the HiTS RRL stars which belong to the Sextans dSph galaxy. 
Using the $M_g$ above, we obtained a distance of $d_{H}=83\pm 4$\,kpc for this galaxy which agrees well with literature values \citep[e.g.,][]{Lee09}.

From the list of distances obtained for the RRLs detected in the HiTS data we closely analyzed stars beyond $100$\,kpc with special attention to potential close pairs or groups on the sky, i.e., stars with small angular separation as well as similar distances. 
A pair and a triplet were obvious in the data, in both cases with angular distances smaller than one degree and a heliocentric distance difference no larger than $10$\,kpc; one of these groups seemed to coincide with the position and published distance of the Leo~V dSph. 
The stars are identified as HiTS113057+021330, HiTS113105+021319 and HiTS113107+021302 (hereafter HiTS113057, HiTS113105 and HiTS113107).
Figure~\ref{fig:lc} shows the light curves of the three RRLs. No search for variable stars in this galaxy has been reported to date, and thus these are the first RRLs detected in this ultra-faint galaxy.

\begin{figure*}
\includegraphics[scale=0.85]{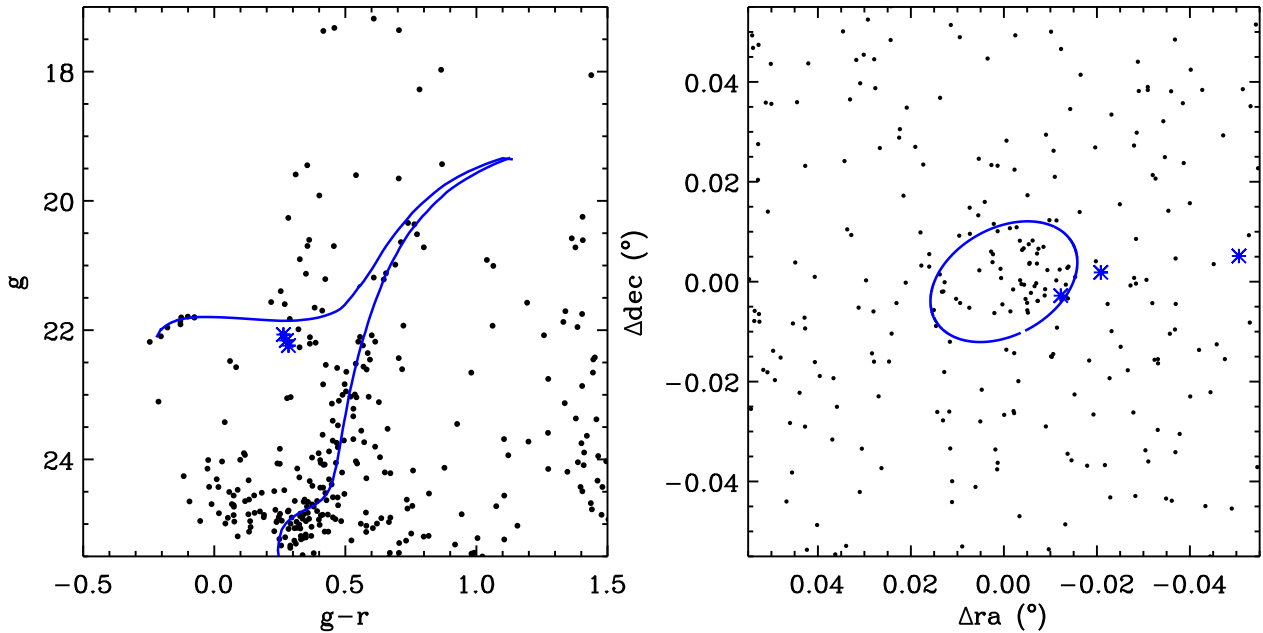}
\caption{{\it left panel:} Color-magnitude diagram of the inner five arcmin of Leo~V. 
The blue stars mark the position of the three RRLs. Overplotted is a $13$\,Gyr old isochrone with [Fe/H]$=-2.2$, visually matched to the BHB stars at a distance of $175.4$\,kpc, consistent with the mean distance of $173\pm5$\,kpc from the RRLs. 
{\it right panel:} Spatial 
distribution of stars near the center of Leo~V. 
The location of the three RRLs is marked. 
The ellipse marks the position of Leo~V's effective radius (Mu\~noz et al. in preparation).}
\label{fig:pos-cmd}
\end{figure*}

Figure~\ref{fig:pos-cmd} ({\it right panel}) shows that, at $0.75$\,arcmin from the center, only HiTS113107 lies within one half-light radius of the dwarf galaxy (with $r_h$=1\,arcmin, according to Mu\~noz et al., in preparation).  
The other two stars lie at $1.26$ and $3.05$\,arcmin (HiTS113057 and HiTS113105 respectively).
Although not as centrally located, they are still close enough to be associated with Leo~V.
Figure~\ref{fig:pos-cmd} ({\it left panel}) shows a color-magnitude diagram (CMD) of Leo~V using data from the Megacam survey by Mu\~noz et al. (in preparation).
The RRLs fall redward of the predominantly blue and sparsely populated blue horizontal branch (BHB), as expected. 
For reference, a $13$\,Gyr old, [Fe/H]$=-2.2$ Padova isochrone \citep{girardi04a, bertelli08a} was visually matched to the blue horizontal branch (BHB) at a distance of $175.4$\,kpc. We note that in our single-epoch CMD, the three RRLs are located below the BHB, at magnitudes of $g=22.07, 22.24$ and $22.16$, respectively, consistent given their amplitudes with having been observed at fainter phases in their light curves (see Fig.~\ref{fig:lc}). 
Table \ref{tab:RRL} summarizes the main properties of the triplet. 

We detected another close group, this time of two RRLs, in addition to Leo~V. 
This pair matched the position of the Leo~IV dwarf galaxy. 
In fact, these two RRLs had been previously discovered by \citet{Moretti2009}. 
These authors identified three RRLs in the Leo~IV region. 
It is worth noting that in Medina et al. (in preparation) we estimated the completeness for detecting RRL at these magnitudes to be $\sim60$\%, which is consistent with the fact that we detected only two of the three RRLs in \citet{Moretti2009}.

We looked for other RRL in our sample that lie spatially close to Leo~V but at closer distances, with the purpose of checking whether possible anomalous Cepheids have been misclassified as RRLs. 
We also looked in the region bridging Leo~IV and Leo~V considering that there is a potential association between the two ultra-faint systems \citep{deJong2010, Blana2012, Jin2012}.
We did not find any other RRL within a radius of $15\arcmin$ from the center of Leo~V, or connecting both dwarf galaxies. 
The closest star was HiTS113107+023025, at $17\farcm 4$ but $\sim 3$ magnitudes brighter than the HB of the galaxy and hence too bright for being an anomalous Cepheid in the galaxy. 
Also, it is located in the opposite direction to Leo~IV, and thus, not in the possible bridge. 
Stars lying close to the three RRL in the CMD of Leo~V shown in Figure~\ref{fig:pos-cmd} ({\it left panel}) were also inspected, resulting in no additional RRL candidates.

We estimated the distance to the Leo~V RRLs in two ways.
As mentioned above, using the mean metallicity of the Galactic halo ([Fe/H]$=-1.6$) as a representative value for our sample, we found a mean $g-$band absolute magnitude of $<M_{g}>=0.69$.
With this value we obtain a mean distance to Leo~V of $163 \pm 4$\,kpc.  
However, it is known that Leo~V is significantly more metal-poor than this value. 
For this reason, a more appropriate estimation should assume a lower metallicity for the RRLs. 
To address this issue, we re-estimate the mean absolute magnitude of our RRLs using as reference the Leo~IV stars in common with \citet{Moretti2009}, and assuming their derived distance of $154\pm5$\,kpc.
Leo~IV has a metallicity of [Fe/H]$=-2.31$ \citep{Simon2007}, which is very close to the recent value of [Fe/H]$=-2.48$ found for Leo~V by \citet{Collins2016}. 
In this case, we derived $<M_g>=0.57\pm 0.07$, which results in a mean heliocentric distance to Leo~V  of $173\pm5$\,kpc.
The individual values for the distances are shown in Table~\ref{tab:RRL}.

\section{Discussion and Conclusions}
    
We have used data from the High cadence Transient Survey in the $g-$band and identified the first three RR Lyrae stars known to date in the Leo~V ultra-faint galaxy. 
From the shape and properties of their light curves, we classified these three RRL, HiTS113057+021330, HiTS113105+021319 and HiTS113107+021302, as fundamental mode pulsators ($ab-$type RRL). 
The periods found for them are $0.6453$, $0.6573$ and $0.6451$\,days, respectively, and the amplitudes, according to fitted models, are $0.72$, $1.34$ and $0.99$\,mag. 

Globular clusters in the MW separate in two groups, \citep[the Oosterhoff, Oo, groups;][]{Oosterhoff1939}, based on both the mean period of their RR Lyrae stars and the proportion between $ab-$type and $c-$type stars. 
Dwarf spheroidal satellites of the MW do not show such dichotomy and many of them have been classified as Oo-Intermediate systems \citep{catelan09,catelan15}. 
Most of the ultra-faint dwarf satellites, on the other hand, despite their low number of stars, have been classified as Oo-II systems \citep[][and references therein]{clementini14}, and Leo~V seems to follow that behavior based on the mean period of $0.65$\,days for the three RRLs.
Unfortunately, the lack of detected $c-$type RRL in this system does not allow verification of this classification based on the proportion between both type of stars.
The properties of the RR Lyrae stars in Leo~V supports the view that building blocks like the ultra-faint galaxies contribute to the Oo-II tail of the MW's halo population, but more massive galaxies, with mostly Oo-I population, are needed to reproduce the present day population of RRLs in the halo \citep{zinn14,fiorentino15,Vivas2016}.
        
In the context of the specific number of RRL in dSphs as a function of magnitude (see Figure 3 from \citealt{Baker2015}), we find that Leo~V is broadly consistent with what has been seen in other ultra faints \citep{Vivas2016}.
However, due to the large dispersion in this correlation, no further conclusions can be drawn just from this additional data.

To derive the distance to Leo~IV we anchored our measured $g-$band magnitudes to a known distance to thus calibrate $<M_g>$. 
In particular, we used data from the low metallicity ultra-faint system Leo~IV, similar in metallicity to Leo~V.
Anchoring our measurements to the RRL stars in Leo~IV \citep{Moretti2009} we obtained a mean heliocentric distance to Leo~V of $173\pm5$\,kpc. 
While consistent within the uncertainties, our value lies on the low side of previously published values.
In their discovery paper, \citet{belokurov08a} used data from the $2.5-$m INT telescope and estimated a distance of $180\pm10$\,kpc to Leo~V. 
\citet{deJong2010} obtained deep photometry of the Leo~IV and V pair with the Calar Alto $3.5-$m telescope and determined a heliocentric distance of $175\pm9$\,kpc.
\citet{Sand2012}, on the other hand, based on images taken with the Clay Magellan telescope, derived a much larger distance of $196 \pm 15$\,kpc. 
The relatively large uncertainties associated to all these measurements are understandable given the sparsely populated Leo~V's BHB. These stars are commonly used as distance indicators and do not require time-series observations, but they lack precision compared to RRLs estimations when the BHB is poorly populated and not well defined, as it is the case for the literature data for Leo~V.

\citet{Baker2015} argued that groups of two or more closely spaced RRLs in the halo can reveal the presence of a Galactic satellite beyond $50$\,kpc. 
The serendipitous discovery of three RRLs beyond $100$\,kpc coinciding with the position of the Leo~V ultra-faint dwarf is a proof-of-concept for their proposal and opens up the exciting possibility of searching for distant ultra-low luminosity, low surface brightness Milky Way satellites. 
This possibility becomes even more relevant when looking ahead to projects like the Large Synoptic Survey Telescope \citep[LSST;][]{LSST2009}; the traditional method for detecting ultra-faint systems based on detecting their major CMD sequences will suffer from significant contamination at the faint end, especially arising from unresolved galaxies, a problem that progressively worsens as we explore the outer regions of the Milky Way halo.
Since RRLs lie in a region of the CMD less contaminated by foreground sources, and particularly due to their identification as pulsational sources, the use of these variables as lightposts for ultra-faint systems will be with no doubt of much valuable help in the efforts to obtain a complete census of  Galactic satellites up to distances of $\sim 400$\,kpc \citep{ivezic08, Oluseyi2012}.

\acknowledgments{
We thank an anonymous referee for his/her careful reading of this article that helped improve the
manuscript.
G.M. and F.F. acknowledges support from the Ministry of Economy, Development, and Tourism's Millennium Science Initiative through grant IC120009, awarded to The Millennium Institute of Astrophysics (MAS), and from Conicyt through the Fondecyt Initiation into Research project No. 11130228. G.M. acknowledges CONICYT-PCHA/Mag\'isterNacional/2016-22162353. R.~R.~M.~acknowledges partial support from BASAL Project PFB-$06$ as well as FONDECYT project N$^{\circ}1170364$.
F.F. acknowledges support from BASAL Project PFB--03 and through the Programme of International Cooperation project DPI20140090. L.G. was supported in part by the US National Science Foundation under Grant AST-1311862. SGG also acknowledges support from the Ministry of Economy, Development, and Tourism’s Millennium Science Initiative through grant IC120009, awarded to The Millennium Institute of Astrophysics (MAS). We acknowledge support from Conicyt through the infrastructure Quimal project No. 140003. Powered@NLHPC: this research was partially supported by the supercomputing infrastructure of the NLHPC (ECM-02). This project used data obtained with the Dark Energy Camera (DECam), which was constructed by the Dark Energy Survey (DES) collaboration. Funding for the DES Projects has been provided by the DOE and NSF (USA), MISE (Spain), STFC (UK), HEFCE (UK). NCSA (UIUC), KICP (U. Chicago), CCAPP (Ohio State), MIFPA (Texas A\&M), CNPQ, FAPERJ, FINEP (Brazil), MINECO (Spain), DFG (Germany) and the collaborating institutions in the Dark Energy Survey, which are Argonne Lab, UC Santa Cruz, University of Cambridge, CIEMAT-Madrid, University of Chicago, University College London, DES-Brazil Consortium, University of Edinburgh, ETH Zurich, Fermilab, University of Illinois, ICE (IEEC-CSIC), IFAE Barcelona, Lawrence Berkeley Lab, LMU Munchen and the associated Excellence Cluster Universe, University of Michigan, NOAO, University of Nottingham, Ohio State University, University of Pennsylvania, University of Portsmouth, SLAC National Lab, Stanford University, University of Sussex, and Texas A\&M University.
}

\bibliographystyle{yahapj}

\clearpage

\end{document}